# Disentangling Spin and Orbital Contributions in the Search for Altermagnetic Terahertz Conversion in $RuO_2$

Jiahao Liu[1,2,§], Yuxiao Mo[1,2,§], Tongyang Guan[1,2], Wenhui Fang[3], Yongwei Cui[1,2], Wentao Qin[1,2], Zhe Yuan[4], Lishu Zhang[3,*], Zhensheng Tao[1,2,*], Yizheng Wu[1,2,*]

[1] *State Key Laboratory of Surface Physics and Key Laboratory of Micro and Nano Photonic Structures (MOE), Department of Physics, Fudan University, Shanghai 200438, China.*

[2] *Shanghai Key Laboratory of Metasurfaces for Light Manipulation, Fudan University, Shanghai 200438, China.*

[3] *Key Laboratory for Liquid-Solid Structural Evolution and Processing of Materials, Ministry of Education, Shandong University, Jinan 250061, China*

[4] *State Key Laboratory of Surface Physics and Interdisciplinary Center for Theoretical Physics andInformation Sciences, Fudan University, Shanghai 200438, China.*

[§] These authors contributed equally to this work.

* Corresponding authors: lishu.zhang@sdu.edu.cn; zhenshengtao@fudan.edu.cn; wuyizheng@fudan.edu.cn.

**Abstract**

Rutile $RuO_2$ has emerged as a prominent candidate altermagnet, yet the microscopic origin of its reported charge-angular-momentum conversion remains unsettled. Recent theoretical and experimental advances have highlighted a potentially dominant role of orbital transport, raising a fundamental ambiguity between spin- and orbital-mediated conversion in $RuO_2$. Here, we report the first experimental disentanglement of the spin- and orbital-to-charge conversion channels in $RuO_2$ heterostructures using terahertz emission spectroscopy. We identify Fe as a predominantly spin-current injector, whereas Ni and permalloy predominantly inject orbital currents, revealing a substantial orbital contribution to ultrafast terahertz emission. Under controlled spin-current injection, we find no evidence for inverse altermagnetic spin-splitting effect. Instead, the observed anisotropy is consistent with an anisotropic inverse spin Hall response that is strongly conditioned by the near-interface environment. Under orbital-current injection, the anisotropy is markedly weaker, indicating that orbital-to-charge conversion is largely isotropic in $RuO_2$. Our results establish orbital transport and interface-conditioned spin Hall conversion as key to terahertz emission from $RuO_2$, underscoring the need to disentangle spin, orbital, and interfacial contributions before assigning anisotropy to altermagnetism.

## Introduction

Rutile $RuO_2$ has become one of the most intensively studied candidate altermagnets. In these collinear, compensated antiferromagnets, crystal symmetry permits momentum-dependent spin splitting, allowing ferromagnetic-like spin polarization to coexist with zero net magnetization [1–5]. A key transport consequence is the altermagnetic spin-splitting effect (ASSE), whereby a longitudinal charge current generates a transverse spin-polarized current whose polarization is governed by the Néel vector and crystal orientation [6–9]. Its reciprocal process, the inverse altermagnetic spin-splitting effect (IASSE) [10–12], converts an injected spin current into a transverse charge current, offering a potential probe of the underlying altermagnetic electronic structure. Although numerous experiments have reported signatures consistent with altermagnetism in $RuO_2$ [6,7,9–14], recent transport, neutron-scattering, muon-spin-rotation, and spin- and angle-resolved photoemission measurements have challenged this interpretation, reporting no evidence for long-range antiferromagnetic order or the proposed altermagnetic band structure [15–19].

Recent advances have added an important orbital dimension to this debate. In the calculated altermagnetic phase of $RuO_2$, the orbital Hall effect (OHE) is predicted to exceed the spin Hall effect by two orders of magnitude and carry the opposite sign, so that orbital-to-spin conversion can generate signals that would otherwise appear spin-Hall-like [20]. Theory further predicts a crystal-orientation dependence of orbital responses in *d*-wave altermagnets [21]. Experimentally, recent torque measurements on (101)-oriented $RuO_2$ have provided evidence for the generation of orbital currents [22]. These findings establish orbital angular momentum as an important degree of freedom in $RuO_2$-based transport. By Onsager reciprocity, the inverse orbital Hall effect (IOHE) should therefore provide a corresponding orbital-to-charge conversion channel under angular-momentum injection. Charge-angular-momentum interconversion in $RuO_2$ can thus no longer be interpreted within a spin-only picture.

This unresolved spin-orbital ambiguity is particularly important for spintronic terahertz emission spectroscopy (STES), which has been widely used to search for the IASSE in $RuO_2$. In a typical STES experiment, femtosecond excitation of a ferromagnetic metal (FM) launches ultrafast spin [23–25] and orbital currents [26–28] into an adjacent $RuO_2$ layer. These currents can be converted into transverse charge currents through the inverse spin Hall effect (ISHE) and IOHE, respectively, generating broadband terahertz emission. If present, the IASSE provides an additional spin-to-charge conversion channel with a characteristic anisotropy governed by the Néel vector and the crystal orientation. Crucially, within the existing theoretical framework, the IASSE requires spin-current injection.

Early STES measurements on the $RuO_2$|permalloy (Py) bilayers reported an anisotropic terahertz response after field annealing and attributed it to the IASSE [12]. More recent studies of the $RuO_2$|Py and $RuO_2$|CoFeB heterostructures, however, have suggested that the IASSE contribution is at most minor, with anisotropic ISHE, the electrically anisotropic conductivity (EAC) of $RuO_2$ [29] and substrate birefringence accounting for most of the observed angular anisotropy [30,31]. Yet these analyses generally assumed predominantly spin-current injection and did not explicitly account for orbital-current-induced terahertz emission. It therefore remains unclear not only which angular-momentum channel dominates the terahertz emission, but also which conversion mechanism governs its angular anisotropy.

In this work, we report the first experimental disentanglement of the spin- and orbital-to-charge conversion in $RuO_2$, establishing distinct spin- and orbital-dominant injection regimes. This separation reveals a substantial orbital contribution to ultrafast terahertz emission in Ni- and Py-based $RuO_2$ heterostructures and, crucially, identifies Fe-based structures as a controlled spin-injection platform for assessing the IASSE. Under such well-defined spin-current injection, we find no evidence for the IASSE. Instead, the observed spin-current-induced anisotropy is consistent with an anisotropic inverse spin Hall response that is strongly conditioned by the near-interface environment. Under predominantly orbital-current injection, the anisotropy is markedly reduced, indicating that the dominant orbital-to-charge conversion is comparatively isotropic. These results provide a framework for identifying altermagnetic charge-angular-momentum conversion in systems where spin, orbital, and interfacial responses coexist.

**Distinguishing Spin- and Orbital-Current-Induced Signals in $RuO_2$|FM**

Figure 1a illustrates the experimental principle. Femtosecond excitation of the FM layer injects spin ($\mathbf{J}_S$) and orbital ($\mathbf{J}_L$) currents into the adjacent $RuO_2$, where they are converted into transverse charge currents through the ISHE and IOHE, respectively, generating terahertz emission. The two contributions are distinguished by complementary polarity and thickness signatures. Because the spin and orbital Hall conductivities of $RuO_2$ are predicted to have opposite signs [20], they produce opposite emission polarities, whereas their distinct transport lengths give rise to different dependences on the $RuO_2$ thickness, $d_R$.

Single-crystal $RuO_2(101)$ wedge films were grown on $Al_2O_3(1\bar{1}02)$ substrates using DC magnetron sputtering, with $d_R$ varying continuously across the 6-mm-long sample at a gradient of ~1.33 nm $mm^{-1}$. A uniform 3-nm-thick FM layer of Fe, Ni, or Py was subsequently deposited on the wedge to serve as the spin- or orbital-current source. These ferromagnets provide distinct injection characteristics: laser-excited Fe predominantly injects spin-polarized currents, whereas Ni generates a substantial orbital current in addition to its spin current [26,32,33]. For Py, which has been widely employed as a spin-source in earlier STES studies [12,30], the dominant character of the injected current is determined experimentally below.

The $d_R$-dependent terahertz emission was measured by scanning the laser spot along the wedge-gradient direction and recording the emitted terahertz fields using free-space electro-optic sampling (EOS) [34,35]. The laser was focused to a spot ~100 μm in diameter, with the effective thickness resolution determined by the spot size, local thickness gradient, and translation step. This approach provides dense, high-resolution sampling of $d_R$, enabling the distinct thickness dependences of the spin- and orbital-current-induced terahertz signals to be resolved [26].

As shown in Fig. 1a, we define a laboratory coordinate system with the sample plane lying in the $x$-$y$ plane and the terahertz emission along the $z$ axis. The $x$ and $y$ axes are aligned with the $RuO_2$ [010] and [$10\bar{1}$] crystallographic directions, respectively. The Néel vector ($\mathbf{L}$) of altermagnent $RuO_2$ is expected to be oriented along [001] and therefore lies in the $y$-$z$ plane, at an angle of ~34.7° with respect to the $z$ axis [36]. During measurements, an external magnetic field $H$~ 1000 Oe was applied in the $x$-$y$ plane to orient the magnetization $\mathbf{M}$ of the FM layer, at a field angle of $\theta_H$ relative to the $y$ axis. The $x$- and $y$-polarized components of the emitted terahertz field, $E_x$ and $E_y$, were selected using a terahertz polarizer before the EOS detection. Further details of the sample fabrication and experimental setup are provided in Supplementary Section S1.

In addition to the spin- and orbital-current-induced contributions, the terahertz emission from the $RuO_2$(101)|FM heterostructures contains an $\mathbf{M}$-independent contribution, which we denote $E_{EAC}$. It originates from the deflection of photoinduced charge currents injected from the FM layer by the EAC of $RuO_2$ [29] and is therefore polarized along the $RuO_2$ [$10\bar{1}$] direction, corresponding to the $y$ axis. We separate it by recording two EOS waveforms at opposite field directions, $\theta_H$ and $\theta_H$+180°. Because the spin- and orbital-current-induced signals reverse sign upon reversal of $\mathbf{M}$, whereas $E_{EAC}$ remains unchanged, the antisymmetric and symmetric combinations $E_{MD}(\theta_H) =$

$[E(\theta_{\mathrm{H}}) - E(\theta_{\mathrm{H}} + 180^{\mathrm{o}})]/2$ and $E_{\mathrm{EAC}}(\theta_{\mathrm{H}}) = [E(\theta_{\mathrm{H}}) + E(\theta_{\mathrm{H}} + 180^{\mathrm{o}})]/2$ yield the **M**-dependent (MD) emission and $E_{\mathrm{EAC}}$, respectively. Details of data analysis are provided in Supplementary Section S2.

Figure 1b compares $E_{\mathrm{MD}}$ of $RuO_2$(4)|Fe, $RuO_2$(4)|Ni, and $RuO_2$(4)|Py, where the number in parentheses denotes the $RuO_2$ thickness in nanometres. Remarkably, the Ni- and Py-based heterostructures share the same emission polarity, opposite to that of $RuO_2$(4)|Fe. This polarity contrast suggests distinct dominant angular-momentum channels: spin-current injection dominates the emission from $RuO_2$(4)|Fe, whereas orbital-current injection dominates that from $RuO_2$(4)|Ni and $RuO_2$(4)|Py, as further supported by their $d_{\mathrm{R}}$ dependences below. The terahertz polarity of $RuO_2$(4)|Fe is also opposite to that of Pt|Fe, a benchmark spintronic terahertz emitter, under the same detection configuration (Supplementary Section S3), indicating an opposite sign of the effective spin Hall response in $RuO_2$ and Pt, consistent with recent studies [20].

Figures 2a-c further show the evolution of $E_{\mathrm{MD}}$ as a function of $d_{\mathrm{R}}$ for $RuO_2$|Py(3), $RuO_2$|Ni(3), and $RuO_2$|Fe(3), respectively. The corresponding field amplitudes extracted at $t$=0 are plotted in Figs. 2d-f. For comparison, these panels also display the thickness dependence of the **M**-independent component, $E_{\mathrm{EAC}}$.

The clear polarity contrast persists over the entire thickness range. $E_{\mathrm{MD}}$ remains negative for both $RuO_2$|Ni and $RuO_2$|Py, whereas it remains positive for $RuO_2$|Fe. Moreover, the Ni- and Py-based heterostructures exhibit remarkably similar $d_{\mathrm{R}}$ dependences (Figs. 2d and e), suggesting a common terahertz emission mechanism. In both cases, the terahertz amplitude rises sharply with $d_{\mathrm{R}}$, peaks at $d_{\mathrm{R}}\approx1$ nm, and then gradually decreases. By contrast, $RuO_2$|Fe displays a more complex, non-monotonic evolution characterized by two maxima (Fig. 2f). The first maximum near $d_{\mathrm{R}}\approx1$ nm is followed by a decrease and then a second maximum near $d_{\mathrm{R}}\approx2.5$ nm before the signal decays at larger thicknesses.

Interestingly, the $E_{EAC}$ signals of all samples display a threshold-like behavior: $E_{EAC}$ amplitude does not increase until $d_R$ reaches a critical thickness $d_{Xtal}$. This behavior can be attributed to the fact that the EAC-induced terahertz emission can only emerge when the $RuO_2$ layer is sufficiently thick to establish anisotropic electrical conductivity. Accordingly, we use $d_{Xtal}$ as a qualitative measure of the thickness at which the $RuO_2$ layer develops sufficiently well-defined crystalline order to support anisotropic electrical transport.

**Distinguishing Bulk and Interfacial Conversion Mechanisms**

The sharp rise of $E_{MD}$ at small $d_R$ in all three heterostructures may initially suggest interfacial terahertz conversion mechanisms. Indeed, recent studies have reported an interfacial Rashba-Edelstein effect at the $RuO_2$|yttrium iron garnet (YIG) interface, which was invoked to explain the opposite signal polarity relative to $RuO_2$|Py [37]. To test this possibility, we inserted a 2-nm Cu spacer layer between $RuO_2$ and the FM layer. If the terahertz emission were dominated by interfacial conversion, Cu insertion would be expected to strongly suppress the signal and substantially alter its $d_R$ dependence.

In Figs. 2g-i, we present the $E_{MD}$ and $E_{EAC}$ amplitudes as functions of $d_R$ for the $RuO_2$|Cu(2)|FM heterostructures. For both $RuO_2$|Cu(2)|Ni and $RuO_2$|Cu(2)|Py, Cu insertion does not alter the overall trend of the $E_{MD}$ thickness dependence (Fig. 2g, h), indicating that the inverse orbital Rashba-Edelstein effect (IOREE) is unlikely the dominant conversion mechanism. Both series show a non-negligible terahertz emission at $d_R$=0 nm with and without Cu insertion. In the bare Ni and Py films, this background arises from the magnetic dipole emission [38,39] and the anomalous Hall effect [40], whereas Cu(2)|Ni and Cu(2)|Py additionally contain an IOHE contribution from Cu [27]. After subtracting these backgrounds, the net peak amplitude $\Delta A_1$ is

reduced by 30-40% upon Cu insertion, which can be attributed to the attenuation of injected orbital currents by the Cu layer. The preservation of the characteristic $d_R$ dependence therefore supports bulk-mediated IOHE in $RuO_2$, rather than interfacial IOREE, as the dominant orbital-to-charge conversion mechanism.

$RuO_2$|Cu(2)|Fe exhibits a qualitatively different response. Although the characteristic double-peak structure persists, the low-$d_R$ peak, $\Delta A_2$, is suppressed by >70%, whereas the high-$d_R$ peak, $\Delta A_1$, is reduced only by ~30% (Fig. 2i). The markedly different sensitivities of the two peaks to Cu insertion indicate that $E_{MD}$ from $RuO_2$|Fe comprises two distinct contributions.

The low-$d_R$ component, dominant below $d_R$≈1.5 nm, is strongly sensitive to the $RuO_2$|Fe interface. Although this behavior may initially suggest an interfacial inverse spin Rashba-Edelstein effect (ISREE) [41,42], a model combining interfacial and bulk conversion fails to reproduce the observed double-peak structure (see below and Supplementary Section S4), arguing against ISREE as the dominant origin. Notably, the $d_R$ < 1.5 nm regime coincides with the onset region of $E_{EAC}$, where the $RuO_2$ crystallinity is not yet fully developed. The first peak is therefore more plausibly associated with enhanced spin-to-charge conversion in a structurally or chemically modified near-interface region, potentially involving Ru-Fe-O intermixing.

By contrast, the high-$d_R$ component, dominant above $d_R$≈2.5 nm, is attributed to bulk-mediated spin-to-charge conversion in crystalline $RuO_2$, such as the ISHE and IASSE. Its ~30% reduction upon Cu insertion is consistent with attenuation of the injected spin current together with electrical shunting by the Cu spacer. Importantly, this bulk-mediated conversion can still be highly sensitive to the near-interface crystalline and electronic environment because the relevant ultrafast transport length is only nanometre scale (see below).

After excluding the interfacial Rashba–Edelstein conversion as the dominant emission mechanism, we model the spin- and orbital-current-induced terahertz signals using a bulk-conversion formalism established for ISHE-driven terahertz emission [24,43]. Here, the spin and orbital channels are treated within the same theoretical framework, enabling direct quantitative comparison of the extracted parameters (e.g., diffusion lengths). The spin- and orbital-associated terahertz amplitudes are described by

$$A_1(d_\mathrm{R}) = A_{S(L)} \cdot \frac{d_\mathrm{FM}}{d_\mathrm{FM}+d_\mathrm{R}} \cdot \tanh\left(\frac{d_\mathrm{R}}{2\lambda_{S(L)}}\right) \cdot Z(d_\mathrm{R}), \tag{1}$$

where $A_{S(L)}$ is the spin (orbital) amplitude coefficient, $d_\mathrm{FM}$ is the FM-layer thickness (fixed at 3 nm in this study), and $\lambda_{S(L)}$ is the spin (orbital) diffusion length. $Z(d_\mathrm{R})$ represents $d_\mathrm{R}$-dependent impedance, given by $Z(d_\mathrm{R}) = \frac{Z_0}{n_1+n_2+Z_0\cdot(\sigma_\mathrm{R}\cdot d_\mathrm{R}+\sigma_\mathrm{FM}\cdot d_\mathrm{FM})}$, where $n_1$ and $n_2$ are the refractive indices of the substrate and air, $Z_0$ is the free-space impedance, and $\sigma_\mathrm{FM}$ and $\sigma_\mathrm{R}$ are respectively the terahertz conductivities of the FM and $RuO_2$ layers, determined from terahertz transmission measurements.

For $RuO_2$|Fe and $RuO_2$|Cu(2)|Fe, an additional phenomenological term is introduced to describe the low-$d_\mathrm{R}$, near-interface contribution:

$$A_2(d_\mathrm{R}) = A_\mathrm{int}\, e^{-\frac{(d_\mathrm{R}-d_0)^2}{2w_\mathrm{R}^2}}, \tag{2}$$

where $A_\mathrm{int}$ is the amplitude coefficient, $d_0$ is the peak thickness of the near-interface contribution, and $w_\mathrm{R}$ is its width. The Gaussian form phenomenologically captures the enhanced conversion in the structurally or chemically modified near-interface regime without assigning it to a specific microscopic mechanism. In particular, an explicit ISREE-based contribution fails to reproduce the characteristic double-peak thickness dependence (Supplementary Fig. S4), further arguing against interfacial Rashba-Edelstein conversion as its dominant origin.

The fitting results, shown as solid lines in Figs. 2d-i, reproduce the $E_\mathrm{MD}$ amplitude as a function of $d_R$ for $RuO_2$|FM and their Cu-inserted counterparts. For $RuO_2$|Fe and $RuO_2$|Cu(2)|Fe, the near-interface and bulk-mediated spin-to-charge contributions are further decomposed, as shown by the dashed lines. The fits yield spin diffusion lengths, $\lambda_S$, of ~1.18 nm for $RuO_2$|Fe and ~1.58 nm for $RuO_2$|Cu(2)|Fe. In contrast, the orbital diffusion lengths ($\lambda_L$) extracted from $RuO_2$|Ni and $RuO_2$|Py, as well as their Cu-inserted counterparts, range from 0.21 to 0.41 nm. This substantially shorter $\lambda_L$ indicates that the injected orbital angular momentum relaxes over a much shorter length scale than spin in $RuO_2$, consistent with recent observations in heavy metals [26]. Importantly, both length scales are nanometre-scale, indicating that the bulk-mediated ISHE and IOHE signals predominantly sample a narrow region of $RuO_2$ adjacent to the interface.

For the Ni- and Py-based heterostructures, the $d_R$-dependent fits include only the dominant IOHE contribution, because the weaker spin-current-induced emission is not separately resolved in the thickness-dependent signals. Nevertheless, such a residual spin contribution may become relevant to the angular anisotropy of the terahertz emission (see below).

**Angular Anisotropy of Spin- and Orbital-Current-Induced Terahertz Emission in $RuO_2$**

Having disentangled the spin- and orbital-current contributions to the terahertz emission, we next examine the angular anisotropy under both spin- and orbital-dominant injection. Because the IASSE is a bulk spin-to-charge conversion process that requires spin-current injection, $RuO_2$|Fe and $RuO_2$|Cu(2)|Fe in the bulk-conversion regime ($d_R$>2.5 nm) provide suitable platforms for assessing the IASSE in $RuO_2$. $RuO_2$|Ni heterostructures, in contrast, provide a complementary platform for investigating the angular response under predominantly orbital-current injection.

In Figs. 3a-b, we present the **M**-dependent components of $E_x$ and $E_y$ terahertz fields as a function of the magnetic-field angle $\theta_H$ for $RuO_2$(4)|Fe and $RuO_2$(4)|Ni, respectively, with the corresponding peak-to-peak amplitudes, $V_{pp}$, indicated. Figures 3c-e summarize the angular

dependence of $V_{pp}$ for $RuO_2$(4)|Fe, $RuO_2$(4)|Cu(2)|Fe, and $RuO_2$(4)|Ni, respectively. Although both field components exhibit approximately sinusoidal dependence on $\theta_H$, the signed peak amplitudes of their angular variations differ markedly among the samples. In $RuO_2$(4)|Fe, the peak amplitude of $E_x$ is ~5% larger than that of $E_y$. By contrast, in $RuO_2$(4)|Cu(2)|Fe, the peak amplitude of $E_y$ is ~32% larger than that of $E_x$, corresponding to an anisotropy of opposite sign. $RuO_2$(4)|Ni exhibits an overall terahertz polarity opposite to that of the Fe-based heterostructures. Nevertheless, its signed anisotropy has the same sign as that of $RuO_2$(4)|Cu(2)|Fe, with the signed peak amplitude of $E_y$ ~11% higher (less negative) than that of $E_x$.

To elucidate the origin of the angular anisotropy, we resolve the measured terahertz field into components parallel ($E_\parallel$) and perpendicular ($E_\perp$) to the magnetization **M**. This transformation provides a convenient symmetry basis for separating the isotropic transverse response from additional anisotropic conversion channels. Accurate conversion from ($E_x$, $E_y$) to ($E_\parallel$, $E_\perp$) (inset of Fig. 4a), however, requires two experimental corrections. First, the birefringence of the $Al_2O_3$ $(1\bar{1}02)$ substrate and the anisotropic terahertz conductivity of $RuO_2$ impose an amplitude anisotropy between $E_x$ and $E_y$. For this correction, the terahertz refractive indices of $Al_2O_3$ and conductivities of $RuO_2$ along $x$ and $y$ were independently determined by terahertz transmission spectroscopy. Second, the magnetic anisotropy of Fe causes the magnetization angle $\theta_M$ to deviate slightly from the applied-field angle $\theta_H$. This deviation can be calibrated by $\theta_H$-dependent magneto-optical Kerr effect measurements, ensuring that the projected amplitudes are referenced to the actual magnetization direction. Details of these corrections are provided in Supplementary Section S5.

Figures 4a-c display the corrected $E_\parallel$ and $E_\perp$ amplitudes as functions of $\theta_M$ for $RuO_2$(4)|Fe, $RuO_2$(4)|Cu(2)|Fe and $RuO_2$(4)|Ni, respectively. In $RuO_2$(4)|Fe, $E_\parallel$ is nearly suppressed, whereas

$E_\perp$ remains essentially independent of $\theta_M$, indicating little intrinsic angular anisotropy after the electrodynamic corrections. The apparent $E_x$-$E_y$ anisotropy in Figs. 3c and f therefore arises predominantly from substrate birefringence and the anisotropic terahertz conductivity of $RuO_2$ [30,31]. By contrast, qualitatively different angular responses emerge in $RuO_2$(4)|Cu(2)|Fe and $RuO_2$(4)|Ni (Figs. 4b, c). In both samples, $E_\parallel$ exhibits a pronounced two-fold angular modulation following $E_\parallel \propto \cos\theta_M \sin\theta_M$, whereas $E_\perp$ consists of a $\theta_M$-independent background superimposed on a $\sin^2\theta_M$ modulation.

**Discussion and Conclusion**

The contrasting responses of the Fe-based heterostructures reveal a pronounced sensitivity of the terahertz anisotropy to the near-interface environment of $RuO_2$. Although $RuO_2$|Fe shows essentially no intrinsic angular anisotropy, a pronounced anisotropy emerges after insertion of a 2-nm Cu spacer. Given the short spin diffusion length in $RuO_2$, this contrast suggests that the anisotropic terahertz response probed by STES is strongly weighted towards the near-interface region.

To further examine this interface sensitivity, Fig. 4d shows the Cu-thickness dependence of the terahertz amplitudes resolved along the two in-plane crystallographic directions of $RuO_2$. We denote the components polarized along [010] and $[10\bar{1}]$ as $E_{010}$ and $E_{10\bar{1}}$, respectively. At small $d_{Cu}$, the two components remain nearly isotropic. Their amplitudes initially increase, reaching a maximum near $d_{Cu}\approx 0.3$ nm, and subsequently decrease as spin-current transmission is attenuated and electrical shunting increases. To quantify their relative difference, we define an anisotropy parameter $\eta = (E_{10\bar{1}} - E_{010})/(E_{10\bar{1}} + E_{010})$. As shown in Fig. 4e, the anisotropy begins to develop above $d_{Cu}\approx 0.6$ nm and becomes nearly saturated around $d_{Cu}\approx 2$ nm. Notably, the EAC-induced terahertz emission also exhibits a maximum around $d_{Cu}\approx 2$ nm (Supplementary Section

S6). The coincidence of these characteristic thickness scales further suggests that the anisotropic terahertz response is closely related to Cu-induced changes in the near-interface crystalline and electronic environment of $RuO_2$.

This raises the question of whether the interface-sensitive anisotropy in $RuO_2$|Cu|Fe could originate from the IASSE. Recent muon-spin-rotation [44] and magneto-optical Kerr effect (MOKE) measurements [45] have reported magnetic order confined to the near-surface region of thin-film $RuO_2$, raising the possibility that altermagnetic-related responses, if present, may be enhanced or confined near the interface.

Within the theoretical description of the IASSE, a spin current $\mathbf{J}_S$ injected along $z$ generates a terahertz field $\mathbf{E}^{\mathrm{IASSE}} \propto \gamma_{\mathrm{IASSE}}\, \mathbf{J}_S \times (\mathbf{M} \cdot \hat{\mathbf{L}})\hat{\mathbf{L}}$, where $\gamma_{\mathrm{IASSE}}$ denotes the IASSE conversion efficiency, and $\hat{\mathbf{L}}$ is the unit vector along the Néel vector $\mathbf{L}$. In the present geometry, the components parallel and perpendicular to $\mathbf{M}$ are given by

$$E_{\parallel}^{\mathrm{IASSE}} \propto -\gamma_{\mathrm{IASSE}}\, j_S \sin\theta_{\mathrm{M}} \cos\theta_{\mathrm{M}}$$

$$\text{and } E_{\perp}^{\mathrm{IASSE}} \propto \gamma_{\mathrm{IASSE}}\, j_S \cos^2\theta_{\mathrm{M}}. \quad (3)$$

However, the injected spin current can also undergo spin-to-charge conversion through the ISHE. Owing to the crystal symmetry of $RuO_2$, the ISHE itself can be anisotropic, yielding [19]

$$E_{\parallel}^{\mathrm{ISHE}} \propto -\Delta\sigma \sin\theta_{\mathrm{M}} \cos\theta_{\mathrm{M}}$$

$$\text{and } E_{\perp}^{\mathrm{ISHE}} \propto \Delta\sigma \cos^2\theta_{\mathrm{M}} + \sigma_0 \quad (4)$$

where $\Delta\sigma = (A - C)\sin^2\theta_C$ and $\sigma_0 = Bcos^2\theta_c + Csin^2\theta_C$. Here, $A$, $B$, and C denote the crystallographic spin Hall conductivity (SHC) tensor components, and $\theta_C$ is the angle between [001] and the $RuO_2$(101) surface normal. Their definitions and the detailed tensor transformation are provided in Supplementary Section S7. Importantly, Eqs. (3) and (4) contain the same angular harmonics. The angular anisotropy observed in $RuO_2$(4)|Cu(2)|Fe is therefore consistent with that expected for the IASSE, but is not unique to it. An anisotropic ISHE (AISHE) can produce the same $\theta_M$ dependence.

The thickness-dependent crystallographic polarization provides an additional criterion for assessing its consistency with the IASSE. In the present geometry, the IASSE contribution is expected predominantly along the $RuO_2$ [010] direction. Experimentally, however, the Cu-induced anisotropy arises mainly from the enhancement of $E_{10\bar{1}}$ rather than $E_{010}$ (Fig. 4d). Thus, although the angular dependence in $RuO_2$(4)|Cu(2)|Fe is compatible with the IASSE, its crystallographic polarization is inconsistent with that expected for the IASSE. Together with the pronounced sensitivity to Cu insertion, this result argues against an IASSE origin and instead points to the AISHE whose anisotropy is strongly conditioned by the near-interface crystal and electronic structure of $RuO_2$.

The Ni-based heterostructures further allow us to examine, for the first time, terahertz anisotropy in $RuO_2$ under predominantly orbital-current injection. Two observations are particularly informative. First, although the overall terahertz polarity of $RuO_2$|Ni is opposite to that of the Fe-based heterostructures, the phase and sign of its angular anisotropy are similar to those observed in $RuO_2$(4)|Cu(2)|Fe (Figs. 4b and c). Second, the relative anisotropy is substantially weaker in the Ni-based heterostructures. As shown in Figs. 4e and g, the absolute value of $\eta$ reaches as large as 0.6 in $RuO_2$(4)|Cu|Fe but remains below 0.12 in $RuO_2$(4)|Cu|Ni even for $d_{Cu}$ up to 4 nm. These observations indicate that the conversion channel dominating the total terahertz emission is not necessarily the one governing its angular anisotropy.

Although the relative magnitudes of the spin and orbital currents injected from Ni cannot be independently quantified, the Ni-based data nevertheless impose an important constraint on their respective conversion anisotropies. The total terahertz emission is dominated by the orbital channel, yet its anisotropy has the same phase as the spin-current-driven AISHE in the Fe-based heterostructures and is inconsistent with that expected from the orbital Hall response. If orbital-to-charge conversion possessed an anisotropy comparable to that of spin-to-charge conversion, the dominant orbital contribution would also dominate the anisotropic response. Instead, the net anisotropy follows the subdominant spin channel. This observation therefore indicates that the orbital-to-charge conversion in $RuO_2$ is substantially more isotropic than its spin counterpart.

Our first-principles calculations support this interpretation. We calculate only the even-symmetry components of the spin and orbital Hall conductivity tensors for antiferromagnetic $RuO_2$ with the Néel vector along [001]. When the spin or orbital current is injected along $[\bar{1}0\bar{1}]$ with its polarization along [010] or $[10\bar{1}]$, the calculated orbital Hall conductivity has the opposite sign to, but a much larger magnitude than, the spin Hall conductivity [20]. Moreover, both Hall conductivities have larger magnitudes for polarization along [010] than for polarization along $[10\bar{1}]$. The calculated relative anisotropy is ~39% for the spin Hall response but only ~9% for the orbital Hall response (Supplementary Section S8). Because the spin- and orbital-mediated terahertz signals have opposite polarities, these tensor anisotropies lead to opposite anisotropy phases in the emitted field, consistent with the experimental discrimination between the two channels. Thus, theory and experiment consistently indicate a substantially weaker orbital than spin Hall conversion anisotropy. A quantitative separation of the two contributions and a direct determination of the orbital Hall anisotropy nevertheless remain challenging and will require independent calibration of the injected spin and orbital currents.

In conclusion, we experimentally disentangle spin- and orbital-to-charge conversion in $RuO_2$

heterostructures, identifying distinct spin- and orbital-dominant injection regimes and establishing a substantial orbital contribution to ultrafast terahertz emission. Under controlled spin-current injection, we find no evidence for the IASSE; instead, the observed anisotropy is consistent with a bulk-mediated anisotropic inverse spin Hall response whose magnitude is strongly conditioned by the near-interface crystalline and electronic environment. Under predominantly orbital-current injection, the terahertz emission remains comparatively isotropic, in agreement with our first-principles calculations. Our results show that spin-orbital disentanglement and near-interface sensitivity are essential for identifying the microscopic origin of terahertz signatures attributed to altermagnetic transport.

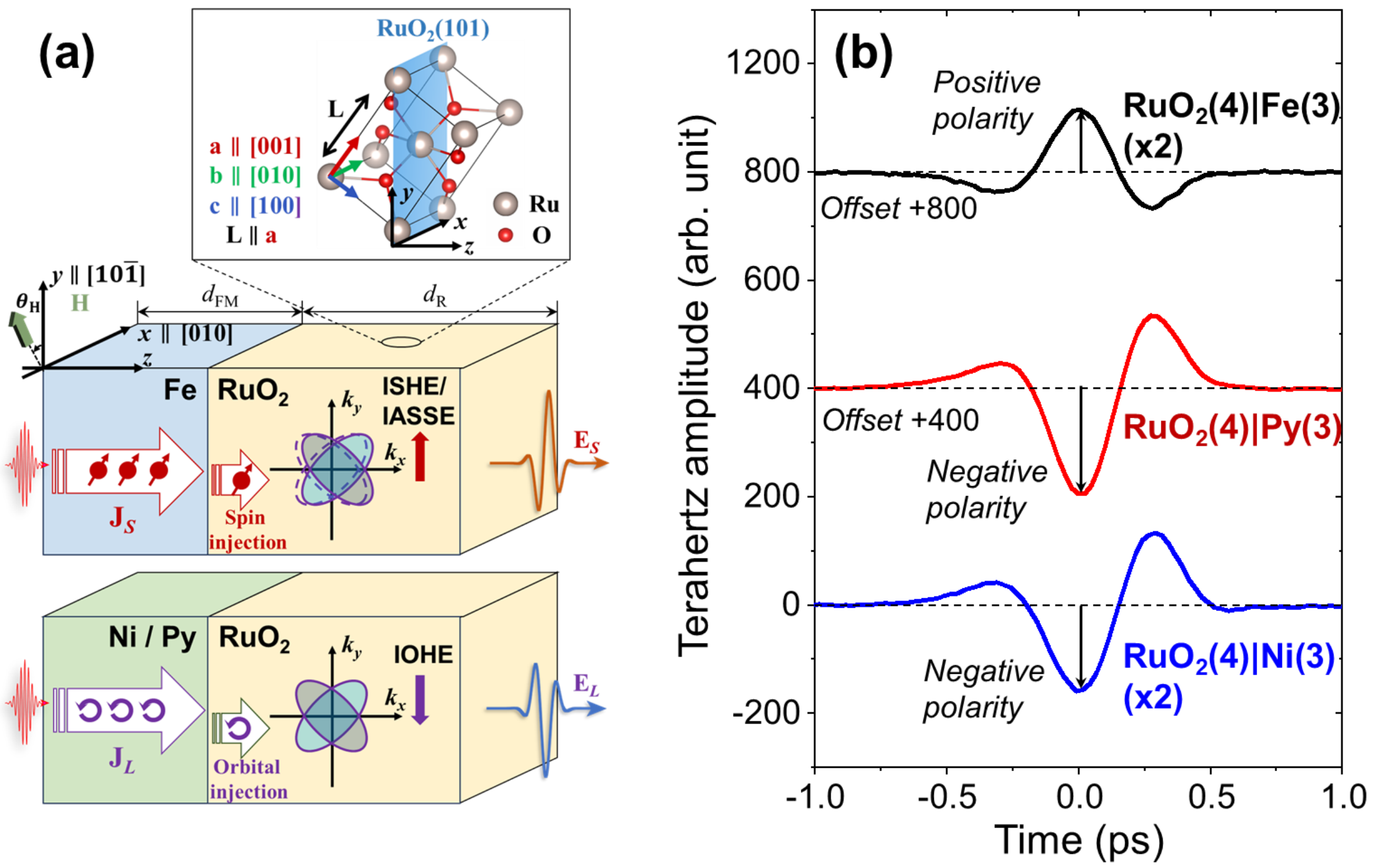

**Figure 1. STES measurements of $RuO_2$|FM heterostructures. (a)** Schematic of the experimental principle for distinguishing spin- and orbital-current contributions to the terahertz emission from $RuO_2$|FM heterostructures with Fe, Ni, and Py as the FM layer. The IASSE is activated only by spin-current injection into $RuO_2$ along *z*, as schematically illustrated by the shifted Fermi surfaces. **Inset:** Unit cell of rutile $RuO_2$. The [001], [010], and [100] crystallographic directions are denoted by **a**, **b**, and **c**, respectively, and **L** indicates the Néel vector. The blue shaded plane denotes the $RuO_2$(101) surface. **(b)** Terahertz signals emitted from $RuO_2$(4)|Fe(3), $RuO_2$(4)|Py(3), and $RuO_2$(4)|Ni(3). For clarity, the Py- and Fe-based traces are vertically offset by +400 and +800, respectively, and the Fe- and Ni-based signals are multiplied by a factor of two. The corresponding terahertz-emission polarities are indicated.

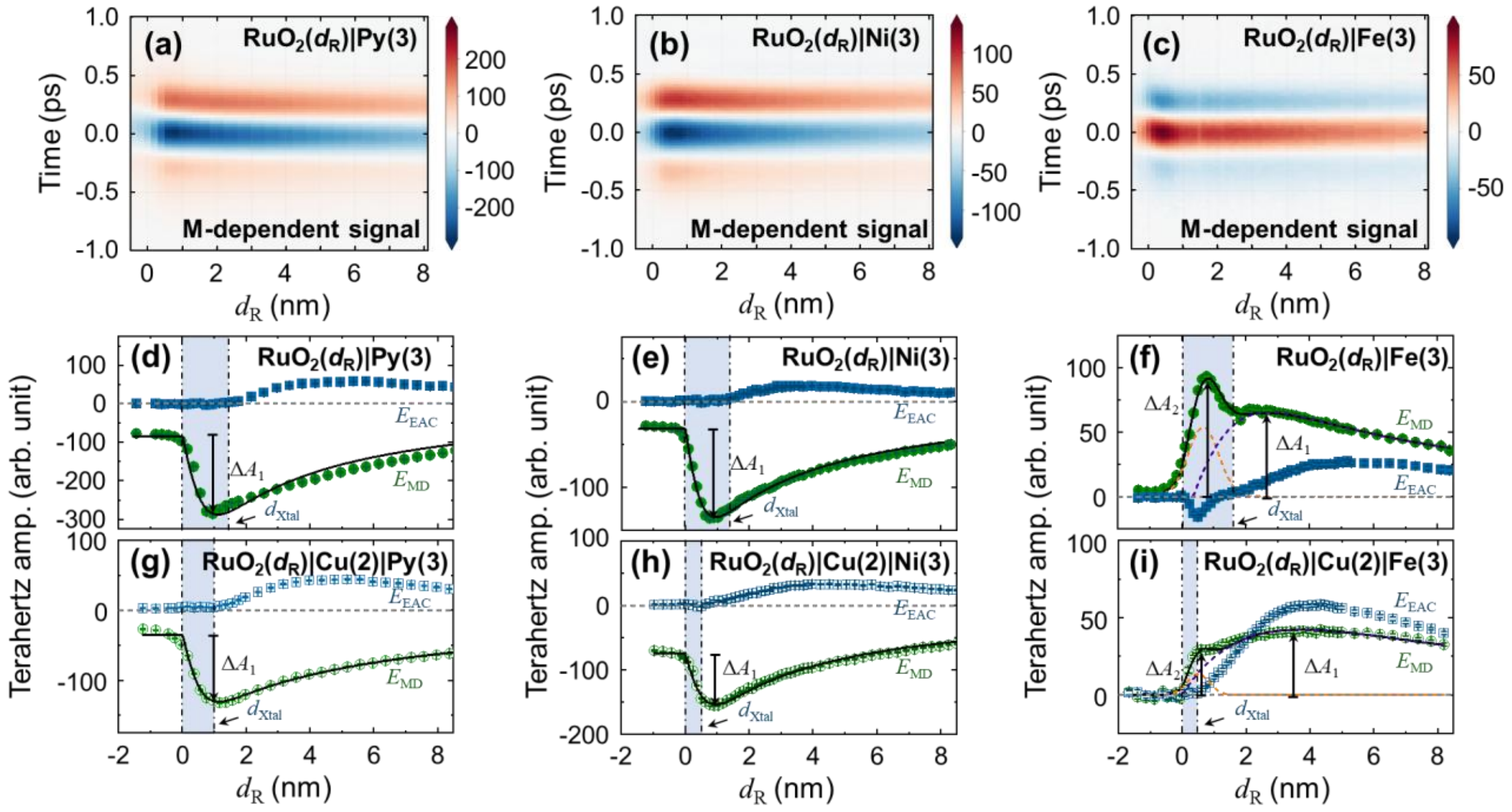

**Figure 2. RuO2-thickness-dependent terahertz emission of RuO2|FM heterostructures. (a-c)** Two-dimensional plots of magnetization-dependent (**M**-dependent) terahertz waveforms as a function of $RuO_2$ thickness, $d_R$, for $RuO_2$|Py(3), $RuO_2$|Ni(3), and $RuO_2$|Fe(3), respectively. **(d)** Terahertz amplitudes extracted at $t$=0 as a function of $d_R$ for $RuO_2$|Py(3). Green circles denote the **M**-dependent component, $E_{MD}$, and blue squares denote the **M**-independent EAC-induced component, $E_{EAC}$. The solid line indicates the model fit. The threshold regime of $E_{EAC}$ is highlighted by the blue shaded region, with the threshold thickness $d_{Xtal}$ indicated. The peak amplitude $\Delta A_1$ is labeled by an arrow. **(e)** Same as **(d)**, but for $RuO_2$|Ni(3). **(f)** Same as **(d)**, but for $RuO_2$|Fe(3). The interface-related peak amplitude $\Delta A_2$ is additionally indicated. Dashed lines represent the interface-related and bulk-related contributions extracted from the fitting. **(g-h)** Same as **(d)**, but for $RuO_2$|Cu(2)|Py(3) and $RuO_2$|Cu(2)|Ni(3), respectively. **(i)** Same as **(f)**, but for $RuO_2$|Cu(2)|Fe(3).

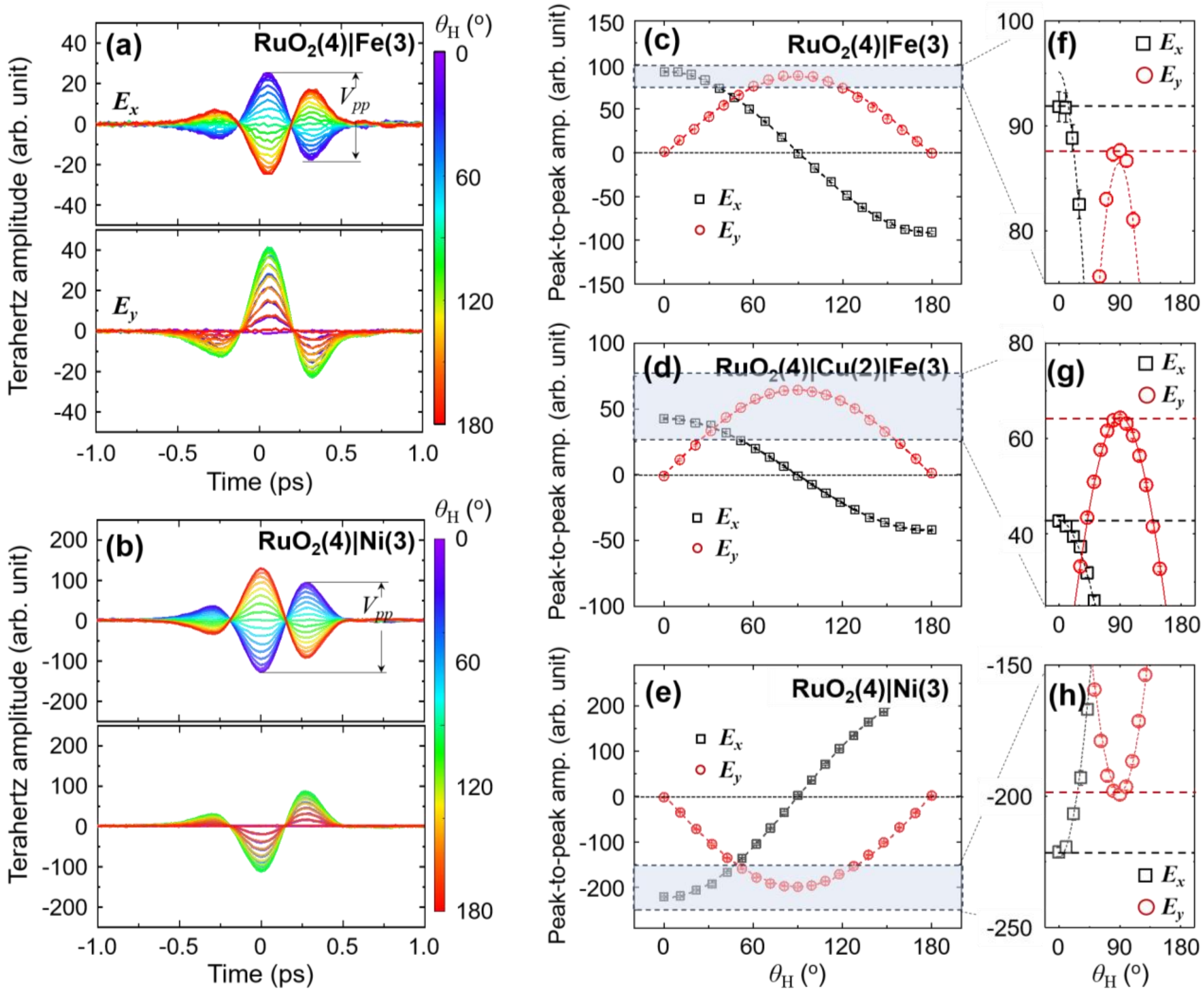

**Figure 3. Angular anisotropy of terahertz emission from RuO2-based heterostructures. (a-b)** $E_x$ and $E_y$ terahertz fields measured under different magnetic-field angles, $\theta_H$, for $RuO_2$(4)|Fe(3) and $RuO_2$(4)| Ni(3), respectively. The peak-to-peak amplitude, $V_{pp}$, is indicated in **(a)**. **(c-e)** Peak-to-peak amplitudes for the $E_x$ and $E_y$ components for $RuO_2$(4)|Fe(3), $RuO_2$(4)|Cu(2)|Fe(3), and $RuO_2$(4)|Ni(3), respectively. Dashed lines show fits to sinusoidal functions. **(f-h)** Zoom-in views of the shaded areas in **(c-e)**, respectively, highlighting the anisotropy between the $E_x$ and $E_y$ peak amplitudes. Black and red dash-dotted lines indicate the peak amplitudes of $E_x$ and $E_y$, respectively.

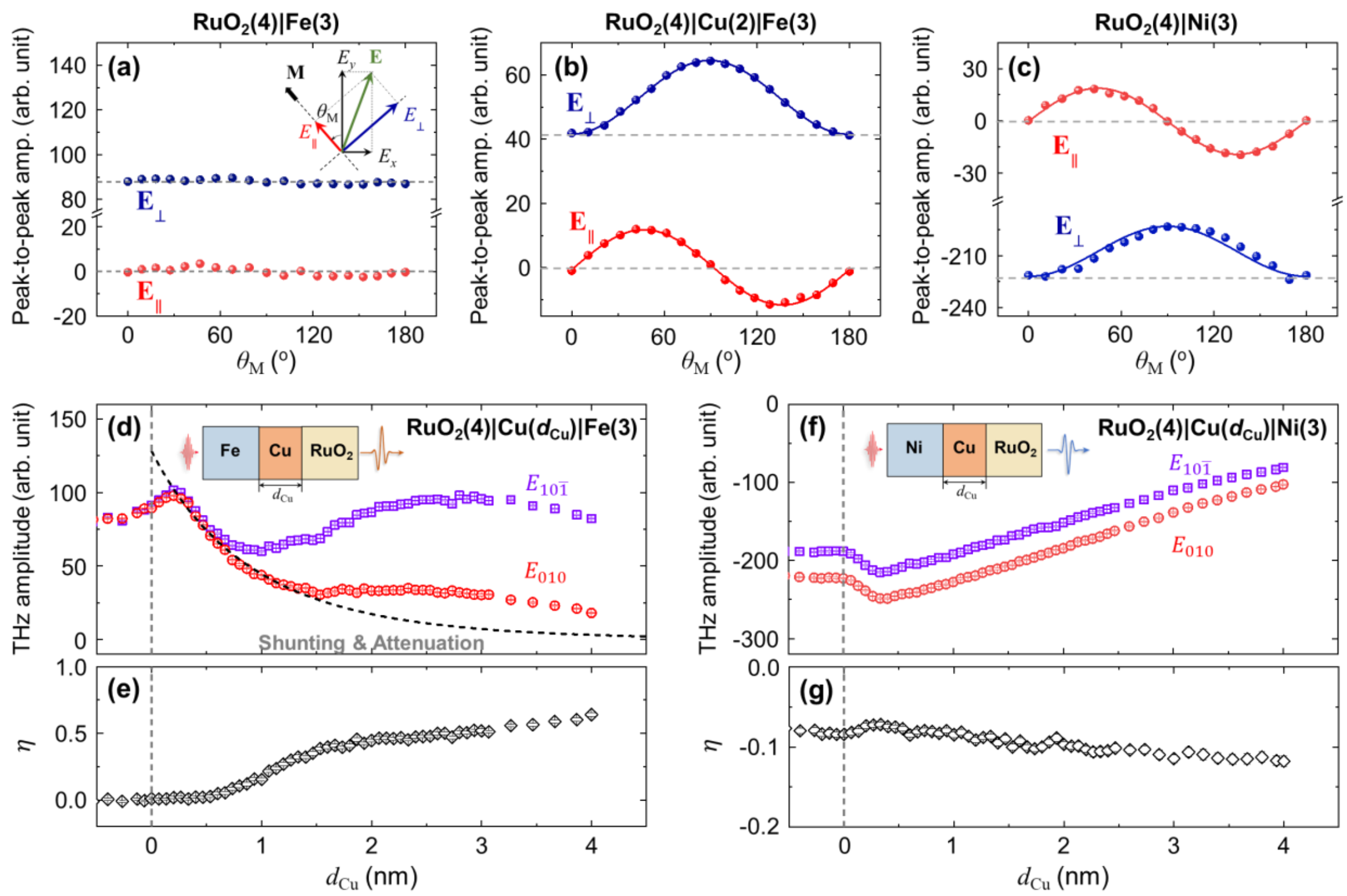

**Figure 4. Angular anisotropy of terahertz emission from Fe- and Ni-based heterostructures. (a-c)** $E_{\parallel}$ and $E_{\perp}$ terahertz amplitudes as functions of the magnetization angle, $\theta_M$, for $RuO_2$(4)|Fe(3), $RuO_2$(4)|Cu(2)|Fe(3), and $RuO_2$(4)|Ni(3), respectively. **Inset** in **(a)**: Schematic illustration of the transformation between the measured $E_x$ and $E_y$ components and the magnetization-referenced $E_{\parallel}$ and $E_{\perp}$ components. Solid lines in **(b)** and **(c)** show fits to the expected angular dependence for the IASSE. **(d)** Terahertz amplitudes polarized along the $RuO_2$ $[10\bar{1}]$ and [010] crystallographic directions for $RuO_2$(4)|Cu|Fe(3), plotted as functions of the Cu-layer thickness, $d_{Cu}$. The dashed line indicates the signal decay expected from attenuation of the injected spin current and increased shunting effect with increasing $d_{Cu}$. **Inset:** Schematic of the experimental geometry. **(e)** Anisotropy parameter $\eta$ as a function of $d_{Cu}$ for $RuO_2$(4)|Cu|Fe(3). **(f-g)** Same as **(d)** and **(e)**, respectively, but for $RuO_2$(4)|Cu|Ni(3).